\documentclass[11pt,a4paper]{article}
\usepackage[utf8x]{inputenc}
\usepackage[T1]{fontenc}
\usepackage[normalem]{ulem}
\usepackage{xcolor}
\usepackage{mathptmx} % Use Times Font
\usepackage{multirow} 
\usepackage{float}
\usepackage{newunicodechar}
\newunicodechar{κ}{\kappa}
\usepackage{makecell}
\usepackage{tabularx}
\usepackage{comment}

\usepackage{etoolbox}
\makeatletter
\patchcmd\@fptop{0pt plus 1fil}{0pt}{}{}
\makeatother

\usepackage{mathtools}

\usepackage[pdftex]{graphicx} % Required for including pictures
\usepackage{subcaption} % For subfigure environment
\usepackage[pdftex,linkcolor=black,pdfborder={0 0 0}]{hyperref} % Format links for pdf
\usepackage{calc} % To reset the counter in the document after title page
\usepackage{enumitem} % Includes lists
\usepackage{color,soul}
\usepackage{amsfonts}  % or
\usepackage{amssymb}

\usepackage[a4paper, lmargin=0.1\paperwidth, rmargin=0.1\paperwidth, tmargin=0.05\paperheight, bmargin=0.1\paperheight]{geometry} %margins

\usepackage[all]{nowidow} % Tries to remove widows
\usepackage[protrusion=true,expansion=true]{microtype} % Improves typography, load after fontpackage is selected

\usepackage{lipsum} % Used for inserting dummy 'Lorem ipsum' text into the template

\usepackage[
backend=biber,
style=numeric,
sorting=ynt
]{biblatex}
\usepackage{bm}
\bibliography{ref}

\newcommand{\review}[1]{\textcolor{red}{review}}

\hypersetup{ 	
pdfsubject = {},
pdftitle = {},
pdfauthor = {}
}

\begin{document}

\title{\LARGE Physics-Enhanced Deep Surrogate for the Phonon Boltzmann Transport Equation}

\author{
    Antonio Varagnolo$^{1}$,
    Giuseppe Romano$^{2}$,
    Raphaël Pestourie$^{1}$\thanks{Corresponding author: rpestourie3@gatech.edu}
}

\date{
    {\small
    $^{1}$School of Computational Science and Engineering,\\
    Georgia Institute of Technology, Atlanta, GA 30332, USA\\[0.5em]
    $^{2}$Institute for Soldier Nanotechnologies,\\ Massachusetts Institute of Technology, 77 Massachusetts Avenue, Cambridge 02139, MA, USA
    }
}
\maketitle

\begin{abstract}
    Designing materials with controlled heat flow at the nano-scale is central to advances in microelectronics, thermoelectrics, and energy-conversion technologies. At these scales, phonon transport follows the Boltzmann Transport Equation (BTE), which captures non-diffusive (ballistic) effects but is too costly to solve repeatedly in inverse-design loops. Existing surrogate approaches trade speed for accuracy: fast macroscopic solvers can overestimate conductivities by hundreds of percent, while recent data-driven operator learners often require thousands of high-fidelity simulations. This creates a need for a fast, data-efficient surrogate that remains reliable across ballistic and diffusive regimes. We introduce a Physics-Enhanced Deep Surrogate (PEDS) that combines a differentiable Fourier solver with a neural generator and couples it with uncertainty-driven active learning. The Fourier solver acts as a physical inductive bias, while the network learns geometry-dependent corrections and a mixing coefficient that interpolates between macroscopic and nano-scale behavior. PEDS reduces training-data requirements by up to ~70\% compared with purely data-driven baselines, achieves roughly 5\% fractional error with only ~300 high-fidelity BTE simulations, and enables efficient design of porous geometries spanning ~12–85 W m$^{-1}$ K$^{-1}$ with average design errors of ~4\%. The learned mixing parameter recovers the ballistic–diffusive transition and improves the out-of-distribution robustness. These results show that embedding simple, differentiable low-fidelity physics dramatically increases the surrogate data-efficiency and interpretability, making repeated PDE-constrained optimization practical for nano-scale thermal-materials design.

\end{abstract}

\section{Introduction}

We introduce a data-efficient and interpretable Physics-Enhanced Deep Surrogate (PEDS) for the phonon Boltzmann Transport Equation (BTE) that makes nano-scale inverse design of thermal materials orders of magnitude faster while preserving accuracy within fabrication error.
PEDS achieves this by embedding a fast Fourier solver inside a neural surrogate (Sec.~\ref{sec:peds}), providing inductive physical bias that reduces training data requirements by up to 75\% compared with purely data-driven models.
Coupled with uncertainty-driven active learning, PEDS requires only~300 high-fidelity BTE simulations to reach 5\% error, making repeated optimization practical~(Sec.~\ref{sec:results}).
This efficiency enables inverse design of porous structures across a wide conductivity range (12–90 W/mK) with average design errors of ~4\%, potentially accelerating thermal management and thermoelectric applications~(Sec.~\ref{sec:efficient-design}).
Moreover, the model’s internal parameters recover the physical transition between ballistic and diffusive regimes, enhancing interpretability and trust in the surrogate model and its designs~(Sec.~\ref{sec:interpret}).

Controlling heat flow at the nano-scale is essential for microelectronics, thermoelectrics, and energy-conversion technologies~\cite{Cahill2003, Vineis2010, Tang2010, wang2016all}. At these length scales the established modeling framework is the phonon Boltzmann Transport Equation (BTE), which resolves the phonon distribution in real and momentum space but is far more expensive to solve than classical diffusive models~\cite{Peierls1929, ziman2001electrons, Cahill2003, Chen2005, cahill2014nanoscale}. This computational cost is especially prohibitive in inverse design settings, where many forward solves for optimization.
% over complex porous geometries are required for optimization~\cite{Evgrafov2009, Evgrafov2011}. 
Inverse design for nano-scale heat transport was first introduced in~\cite{Evgrafov2009}, where optimal material distributions were identified for various problems, such as maximum dissipation and temperature control. Subsequently, Ref.~\cite{romano2022inverse} developed a differentiable phonon BTE solver based on JAX~\cite{Bradbury2018} to design thermal metamaterials with prescribed effective thermal conductivity. This tool, which was merged in~\texttt{OpenBTE}~\cite{openbte}, performed density-based topology optimization~\cite{gersborg2006topology, lundgaard2018density} based on the three-field approach~\cite{SigmundMaute2013} and a novel interpolation technique to map the density of the material into an effective transmission coefficient. 

Despite this advance, a computational bottleneck arises in scenarios where multiple optimization runs need to create multiple target conductivities, because the topology optimization needs to be rerun from scratch for each target, which can become prohibitive due to the cost of solving the BTE.

%WHAT WE DID

%PREVIOUS AI APPROACHES(in general and the one for nano-scale, which is the MFP)
To address the computational costs, data-driven surrogates aim to accelerate the simulation and optimization of typically high-dimensional, design parameters search~\cite{GarciaEsteban2021, karniadakis2021physicsinformed}. In particular, machine learning surrogates hold the promise to learn the mapping from design parameters (i.e. material structure) to a desired low-dimensional target property (i.e. effective thermal conductivity) and dramatically reduce computational expense for the evaluation of the property. In addition, they offer a continuous relaxation of the parameters and the gradient may be computed efficiently via automatic differentiation~\cite{griewank2008evaluating, baydin2018automatic}. However, purely data-driven models come with the cost of the training data and are limited by the curse of dimensionality that require large training sets that grow exponentially with the number of variables~\cite{bishop2006pattern, hastie2009elements}, causing an upfront, potentially large, training cost. Moreover, surrogate models may generalize poorly outside the generated data distribution, being often unreliable at extrapolation compared to interpolation~\cite{raissi2019physics}. Hybrid models from scientific machine learning (SciML) aim at reducing this data need by incorporating domain knowledge inside the model. Scientific and physics-informed approaches (SciML, PIML) have been used to create faster approximations~\cite{karniadakis2021physicsinformed,peurifoy2018nanophotonic,So2020DeepID,Kiarashinejad2020DimRed,kiarashinejad2020knowledge,kiarashinejad2019deep,An2019Objective,An2020Deep}. These include physics-informed networks~\cite{raissi2019physics,lu2021physics}, solver corrections~\cite{chung2024scaled, dresdner2022learning}, and operator learning~\cite{li2021fourier,kovachki2023neural,lu2021learning}. Many of these models follow physical rules by including symmetries~\cite{cohen2016group}, structural patterns~\cite{baddoo2023physics}, or geometric information~\cite{bronstein2017geometric,gao2022physics}, or by directly embedding solvers in the learning process ~\cite{Pestourie_Mroueh_Rackauckas_Das_Johnson}. This approach has enabled major results in protein folding~\cite{jumper2021highly} and weather forecasting~\cite{pathak2022fourcastnet,kochkov2024neural}. Physics-Informed Neural Networks (PINNs) ~\cite{raissi2019physics}  have been extensively applied to diffusive ~\cite{Mishra2020} and ballistic heat transfer~\cite{li2021physicsinformedneuralnetworkssolving, li2022physics, li2023physics, lin2025monte} and to inverse problems in the diffusive ~\cite{Cai2021, Billah2023, Qian2023} and ballistic regime ~\cite{ zhou2025physics}. Focusing on nano-scale transfer, in ~\cite{li2021physicsinformedneuralnetworkssolving, li2022physics, zhou2023physics, lin2025monte}, PINNs are trained minimizing the sum of the residual with respect to the mode-resolved BTE and its boundary conditions at different points accross the domain (data-free ML). Recently, Ref.~\cite{zhou2025physics} proposed an architecture that extends to identifying some of the PDE parameters in an inverse problem setting. However, PINNs are solvers that learn a PDE solution rather than surrogate models that learn the parameterized function of a property. PINNs are typically not parameterized, but previous work introduced one physical parameter to the solver~\cite{li2021physicsinformedneuralnetworkssolving, li2022physics, zhou2023physics, lin2025monte} ––– the characteristic length which is hard to define and compute for complex geometries. In contrast, our surrogate model is geared towards design with twenty-five geometry parameters.
Multi-fidelity (MF) DeepONets~\cite{lu2021learning} learn a parameterized PDE solver from data, integrating low-fidelity simulations with sparse high-fidelity data~\cite{lu2022multifidelity}. This approach improved predictive accuracy and reduced the cost of data generation. However, it required 1000 high-fidelity data, and the low-fidelity model still required 10,000 approximate BTE solves to work, which remained computationally expensive. In contrast, we need only a few hundred high-fidelity data and use the much cheaper macroscopic approximation of Fourier equation inside our model. Note that operator learning techniques like DeepONet reduce to a single neural network (one of our baselines) when used as a parameterized surrogate model~\cite{Pestourie_Mroueh_Rackauckas_Das_Johnson}.

We developed a SciML surrogate for the steady-state Boltzmann Transport Equation based on Physics Enhanced Deep Surrogate (PEDS)~\cite{Pestourie_Mroueh_Rackauckas_Das_Johnson}. PEDS combines a low-fidelity solver, enforcing physical behavior, with a neural network that learns the low-fidelity solver input that makes it accurate for a target property. The cheap solver provides an inductive bias~\cite{Mitchell1980} with a relaxed version of the physics (Sec.~\ref{sec:peds}). In our case, we employ a differentiable Fourier simulator as the low-fidelity approximation of the BTE. To our knowledge, this is the first successful implementation of a multifidelity approach that leverages the diffusion equation for BTE. Although when computing the thermal conductivity it has a fractional error of up to 600\%, the Fourier solver is $\approx$ 2300 times faster to compute (11000 times faster considering its strong batch-parallelism) and its inductive bias accelerates training and improves the generalization in out-of-distribution regions, resulting in three to four orders of magnitude accelerations of PEDS compared to BTE. Solving one high-fidelity BTE takes around 3 minutes on 4 CPUs for our illustrative example, so training costs are dominated by data simulation costs, and our method aims at reducing the data need while retaining accuracy. Combined with active learning~\cite{pestourie2020active}, PEDS needs only 300 data points to achieve a 5.00\% target fractional error (Sec. ~\ref{sec:results}), dominated by fabrication error and sufficient for PDE-constrained optimization purposes (Sec. ~\ref{sec:efficient-design}). The proposed surrogate outperforms a purely data-driven baseline, improving the model test set fractional error of up to 76\% relatively to the data-driven baseline for the same number of training points (Sec. ~\ref{sec:results}).  
% Inspired by~\cite{Pestourie_Mroueh_Rackauckas_Das_Johnson}, we incorporated an heteroskedastic measure of uncertainty. Uncertainty Quantification (UQ) improves the robustness of the model, preventing overconfident but inaccurate prediction especially when designing in regions under-sampled in the training set~\cite{psaros2023uncertainty, abdar2021review}. Finally, it enables further data efficiencies by use of Active Learning (AL). The pairing of PEDS and AL allows substantial computational savings at data generation time compared to the other baselines. At parity of validation set fractional error, active learning strategy allows us to save 70\% (300 training points vs 1000 training points) of generated data points to reach 5\% fractional error. This is equivalent to around 32 hours of computation time, 
Our current inverse design pipeline enables fast PDE evaluations for thermal conductivity design tasks. We achieve an average design error of $\sim$4\% on 8 example design objectives, dominated by the error of the fabrication process. Thanks to the data efficiency of PEDS and Active Learning (AL), we fully amortize the training costs with only four designs.

\section{An Overview of Physics Enhanced Deep Surrogates}
\label{sec:peds}

\begin{figure}[H]
    \centering
    \includegraphics[width=1.0\linewidth]{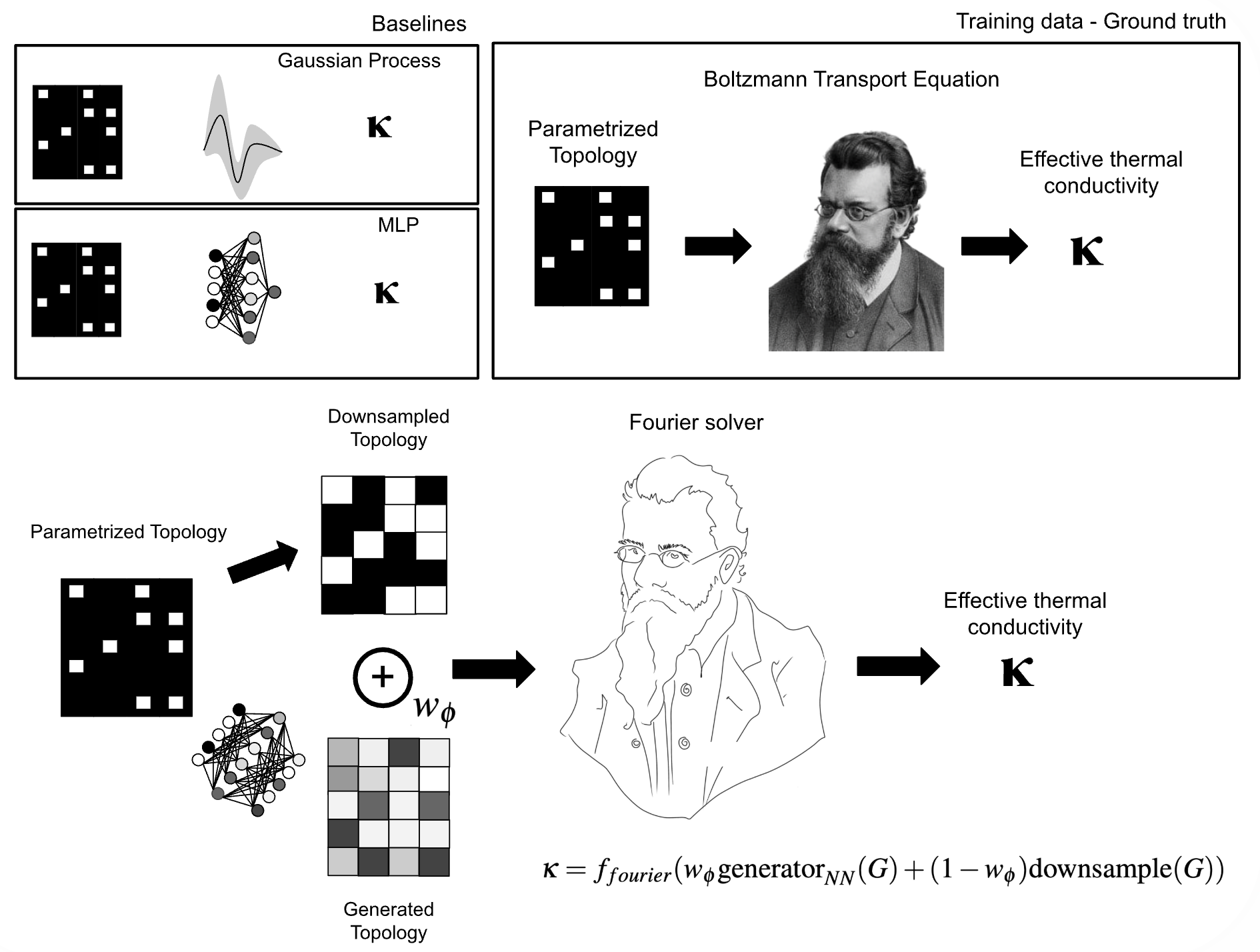}
    \caption{\textbf{PEDS Diagram} The main panel illustrates the PEDS workflow: starting from a vector encoding the parametrized topology $G$, a neural-network generator $\text{generator}_{NN}$ produces a coarser, non-linear transformation of the design space. This generated topology is then combined with a coarsified representation of the original geometry via a linear combination; the mixing coefficient $w_\phi \in [0,1]$ is itself learned as a function of the geometry parameterization. The resulting linear combination is passed to a computationally cheap, low-fidelity version of the physics. In our case this is the Fourier solver (depicted here with a cartoon image of Ludwig Boltzmann) which evaluates the approximate physics and outputs the target macroscopic property. The top right inset highlights how the training dataset is constructed: high-fidelity ground truth labels are produced by running expensive Boltzmann Transport Equation simulations using the OpenBTE software (represented here by a historical photograph of Ludwig Boltzmann, courtesy of the University of Frankfurt). The top left insets represent the considered baselines: a Gaussian Process with a Matern kernel and an MLP.}
    \label{fig:peds}
\end{figure}

\begin{figure}[H]
    \centering
    \includegraphics[width=1\linewidth]{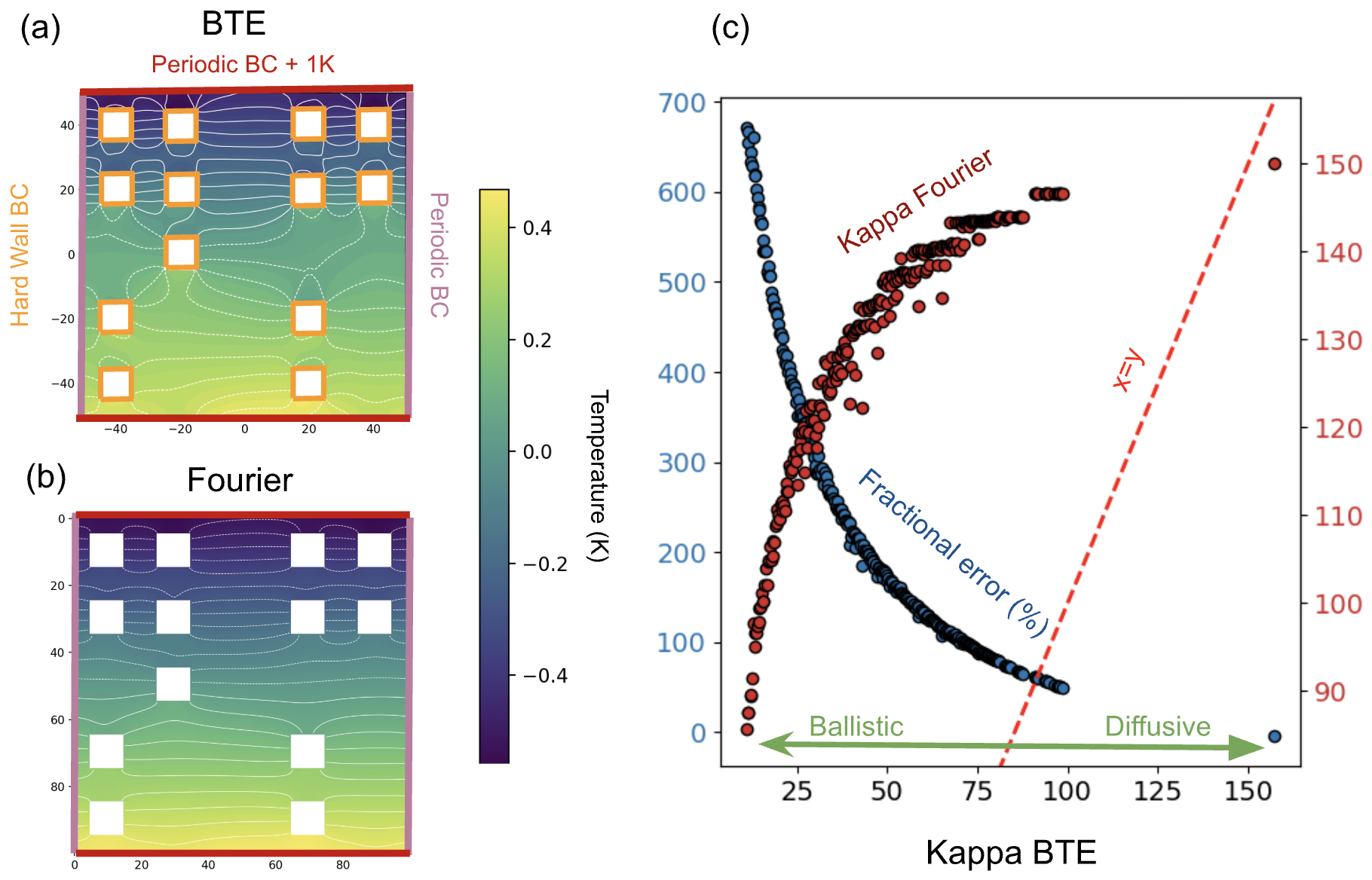}
    \caption{Comparison between the Fourier solver and the BTE. On the left we show the qualitative difference between the temperature fields obtained with (a) the ballistic BTE and (b) the diffusive Fourier model for the same geometry and problem setup. The domain is a square of size 100 nm, containing embedded square pores of size 10 nm, with periodic boundary conditions applied along both the x and y directions, and a temperature difference imposed between top and bottom of the domain. The effective thermal conductivity $\kappa$ obtained by the Fourier temperature field is \textbf{105 W/mK}, while it is \textbf{23 W/mK} for the BTE field. (c) On the right we present a quantitative comparison of the computed effective thermal conductivity for 500 nanoporous designs in our test set.}
    \label{fig:fouriererror}
\end{figure}

PEDS is a scientific machine learning model comprising a computationally cheap low-fidelity solver paired with a neural network generator~\cite{Pestourie_Mroueh_Rackauckas_Das_Johnson}, as in Eq.~\ref{eq:peds} and illustrated in Fig.~\ref{fig:peds}. The neural network applies a nonlinear transformation to the design-parameter space; the transformed parameters are passed to the low-fidelity solver, which approximates the governing physics and returns the target property. The whole architecture is trained end-to-end to predict the target property. PEDS belongs to the input-space representation machine learning field~\cite{koziel2008space, marzban2025inverse} and is similar to neural space mapping~\cite{bakr2000neural} or coarse/fine grid mapping~\cite{feng2019coarse}, with the relevant differences that the output parametrization is not the same as the initial and that in more classic input-space machine learning there is no mixing of the generated input with an input from field knowledge. The low-fidelity solver may be the same numerical method as the high-fidelity PDE solver, run at lower spatial resolution or with higher tolerance, or it may incorporate deliberate simplifications of the physics (for example, by linearizing a nonlinear term). This low-fidelity solver can produce large errors in the target output (in our case 220 \% on average as it can be seen in Figure ~\ref{fig:fouriererror}), but it is orders of magnitude faster than the high-fidelity model while qualitatively preserving at least some of the underlying physics (e.g. the domain boundary conditions). To incorporate further physical bias, the geometry fed to the approximate solver can be a linear combination of the generated topology and the original geometry, possibly coarsified to match the generated topology dimension. The coefficient of the convex linear combination, from now on referred to as the \textit{mixing coefficient}, is learned as a function of the input geometry. The modified geometry can also enforce physically sound biases like symmetries. The neural network weights are learned contextually using backpropagation and the adjoint simulation of the PDE. PEDS has previously shown great improvements in terms of accuracy and data efficiency compared to other deep parametric models and classical surrogates in similar tasks, including surrogate modeling for diffusion and diffusion-reaction equations, and for the more complex Maxwell's equations~\cite{Pestourie_Mroueh_Rackauckas_Das_Johnson}. PEDS is defined as

 \begin{equation}
    \kappa \approx f_{fourier} \left( w_\phi \text{generator}_{NN}(G) + (1 - w_\phi) \text{downsample} (G)\right), 
    \label{eq:peds}
 \end{equation}

\noindent where $G$ parameterizes the surrogate model input geometry, $f_{fourier}$ is the low-fidelity solver, $w_\phi$ the mixing coefficient, $\text{generator}_{NN}$ the neural generator of the solver input, and $\text{downsample}$ the solver input from field knowledge.
PEDS fits greatly the Boltzmann Transport Equation, as the mixing coefficient $w_\phi$ adapts predictions between macro- and nano-scale effective conductivities, naturally capturing the smooth transition from diffusive to ballistic transport. Moreover, there is wide literature converging to a clear choice for the solver to approximate the physics: the Fourier equation, also known as heat conduction equation or Poisson equation. The Fourier Equation is the macro-scale approximation of the BTE and it can be derived through simplification of the BTE. In Fig. ~\ref{fig:fouriererror}a and ~\ref{fig:fouriererror}b the temperature fields of BTE and Fourier are shown for a representative geometry. Looking at the contour lines, it is easy to appreciate the difference: in the BTE solution, the temperature field exhibits clear non-diffusive effects, with strongly distorted isothermal contours around the pores that highlight the influence of boundary scattering. In contrast, the Fourier solver produces a smooth field with symmetric and evenly spaced isothermal contours. The effective thermal conductivity obtained from BTE for this case is substantially damped: 23 $W/mK$ against the Fourier solution $105 W/mK$. In Fig. ~\ref{fig:fouriererror}c, a comparison of the effective thermal conductivity obtained through BTE and Fourier is showcased. The Fourier field systematically overestimates the thermal conductivity when compared with the BTE results (see the line x = y for reference). The overestimation is most severe for designs where the BTE conductivity is small, corresponding to cases dominated by ballistic transport, with fractional errors exceeding 700\% in the most extreme examples. When the average conductivity approaches the bulk conductivity (i.e. assuming uniformity of the material and absence of pores), the BTE scalar temperature field approaches the Fourier field. The Fourier Equation meets the inclusion criteria defined in~\cite{Pestourie_Mroueh_Rackauckas_Das_Johnson}  , since uniform geometries encompass the full range of effective conductivities described by the BTE.\\

\noindent The neural network parameters and mixing coefficient are learned jointly in an end-to-end manner, making a differentiable low-fidelity solver essential. To efficiently compute the gradients of the PDE solution with respect to multiple design parameters, we exploit the adjoint simulation, also known as reverse-mode automatic differentiation~\cite{baydin2018automatic} or backpropagation, depending on the reader's background. Instead of using system inversion or forward numerical differentiation, gradients are obtained by solving the adjoint equation, an auxiliary PDE that captures efficiently the solution’s sensitivity to design parameters. Obtaining the gradients amounts only to solving another system similar to the forward system. For self-adjoint operators, like Fourier with periodic boundary conditions, the adjoint system is identical to the original, and further computational efficiencies can be achieved depending on the re-usability of the forward solution strategy. Training and inference were performed on Intel Core i5 quad-core processors (2.3 GHz, 2017 generation) using CPU execution only.

\subsection{Uncertainty Quantification with Deep Ensembles and Heteroskedastic Variance}\label{sec:NLL}
Uncertainty Quantification (UQ) is essential in surrogate modeling for scientific applications~\cite{psaros2023uncertainty}, especially when high-fidelity simulations are computationally expensive and data is limited~\cite{abdar2021review}. We adopt a heteroskedastic Gaussian surrogate model, where the mean $\kappa_\theta(x) \in \mathbb{R}^N$ is predicted by an ensemble of PEDS~\cite{lakshminarayanan2017simple} and the input-dependent log-variance $(\log \sigma^2)_\theta(x)$ is predicted by a vanilla neural network. The model is trained using a Gaussian negative log-likelihood loss~\cite{bishop2006pattern, lakshminarayanan2017simple}: $\mathcal{L}_{\text{NLL}} = \sum_i \beta \cdot (\log \sigma^2)_\theta(x_i) + (1 - \beta) \cdot \frac{(\kappa(x_i) - \kappa_\theta(x_i))^2}{\exp((\log \sigma^2)_\theta(x_i))}$, where the parameter $\beta \in [0, 1]$ controls the trade-off between penalizing overconfident predictions and fitting the data. We train an ensemble of $M$ independently initialized surrogate models to further capture uncertainty arising from model variability, especially from the optimization of their parametric parts. The ensemble prediction is given by the empirical mean $\hat{\kappa}(x) = \frac{1}{M} \sum_{i=1}^{M} \kappa_i(x)$, and the total predictive variance combines the predicted variance and the variance of the predictions as $\hat{\sigma}^2(x) = \frac{1}{M} \sum_{i=1}^{M} \sigma_i^2(x) + \frac{1}{M} \sum_{i=1}^{M} (\kappa_i(x) - \hat{\kappa}(x))^2$. This approach robustly predicts epistemic uncertainty, supports active learning by prioritizing uncertain samples, and facilitates stochastic optimization by taking into account the confidence in the surrogate’s output rather than relying solely on point estimates.

\subsection{Active Learning}
PEDS has previously shown great data savings and better performance on challenging designs when paired with an Active Learning framework~\cite{Pestourie_Mroueh_Rackauckas_Das_Johnson}. We implemented an uncertainty-driven active learning pipeline using our uncertainty measure following Ref.~\cite{pestourie2020active}. We initialize a dataset with N training points, we train our surrogate and then propose M new geometries, compute their uncertainty and only compute and add to the dataset the $K\ll M$ most uncertain points. This is in contrast with other sample proposal strategies that are based upon the diversity of the samples or a mixture of diversity and uncertainty~\cite{settles2012active, shui2020deep}. In our case the measure of uncertainty is the variance as computed above. For active learning, unlike complex data assimilation problems requiring detailed distribution estimates, a coarse approximation of uncertainty is sufficient~\cite{abdar2021review}.

%HOW THE PAPER IS STRUCTURED

\section{Design of Porous-structure with desired effective Thermal Conductivity}

\label{sec:design}

% The ability to engineer the thermal properties of nanostructured materials is crucial for a variety of applications, from heat management in microelectronics to the design of efficient thermoelectrics~\cite{Cahill2003, Vineis2010, Tang2010, wang2016all}. 
At the nano-scale, heat conduction diverges from the classical Fourier Law~\cite{Chen2005, Chen2021} and assumes the form of a ballistic process rather than a diffusive one. This is because the mean free path (MFP) of heat-carrying quasi-particles in semiconductors, i.e. \textit{phonons}, is comparable to the characteristic dimensions of the material. This phenomenon, often referred to as \textit{phonon-size effect}~\cite{Chen2005, Tang2010}, leads, in porous material, to a suppressed effective thermal conductivity compared to the one obtained with a standard Fourier solver~\cite{Song2004, Hochbaum2008, Lee2015} (Fig.~\ref{fig:fouriererror}). These effects hold great promise across multiple engineering domains, shifting the focus from bulk materials to nanostructures and their inverse design. 

In recent work, direct topology‐optimization of porous nanostructures has been attempted using the phonon BTE. For example, Ref.~\cite{Wei2020} employed a genetic‐algorithm search with a gray (single‐MFP) BTE model for nanoporous graphene and found disordered pore patterns that increase thermal conductivity. However, gray (single mean-free-path) models neglect the full phonon spectrum and generally overestimate conductivity. In fact, gray-model solutions can qualitatively differ from full-spectrum solutions: for instance, Ref.~\cite{iheduru2023comparison} shows that a gray BTE systematically predicts a higher effective conductivity than a mode-resolved BTE for the same nanofilm. Thus, designs based on single‐MFP models may yield very different results than those from a full phonon-spectrum BTE. The development of differentiable BTE frameworks is an active research area. The first implementation of such a framework in the context of inverse design (Ref.~\cite{romano2022inverse}) focused on the single-MFP model; only recently, a mode-resolved differentiable BTE solver (based on the relaxation-time approximation) was developed~\cite{shang2025jax}; however, this framework did not include the interpolation model needed for inverse design (Ref.~\cite{romano2022inverse}). 

With the surrogate framework in place, we now focus on the inverse design of two‑dimensional porous geometries whose effective thermal conductivity can be tuned over a broad range. Our surrogate directly outputs an accurate estimate of the steady‑state BTE conductivity and we can therefore drive a vast array of off-the-shelf optimizers to hit any target conductivity. In the following, we demonstrate this approach on a simple geometry and with six target conductivities ranging from 12 to 85 W/mK.  

\subsection{Problem Statement}
We consider a 2D domain $\Omega = [-L/2, L/2] \times [-L/2, L/2]$ where $L$ = 100 nm illustrated in Fig.~\ref{fig:fouriererror}. We model nano-scale heat transport via the Boltzmann transport equation under the relaxation time approximation, which in the pseudo-temperature formulation~\cite{romano2015heat}, reads
\begin{equation}\label{eq:BTE}
     - v_{\mu} \cdot \nabla T_{\mu}(\mathbf{r}) = \dfrac{T_{\mu}(\mathbf{r}) - \bar T(\mathbf{r})}{\tau_{\mu}} 
     \quad \quad \quad \quad \quad \quad \quad \mathbf{r} \in \Omega,
\end{equation}
where $T_\mu(\mathbf{r})$ is the phonon distribution for mode $\mu$, which collectively describes wave vector and polarization. The term $\mathbf{v}_\mu$ is the mode-resolved group velocity, and $\tau_\mu$ denotes the scattering time for mode $\mu$, which includes three-phonon as well as naturally occurring isotope scattering~\cite{ziman2001electrons}. The term $\bar T(\mathbf{r})$ is the pseudo-temperature~\cite{hao2009frequency}, which is the quantity plotted in the examples. Its expression is~\cite{romano2021efficient}
\begin{equation}
\label{eq:temp}
    \bar T(\mathbf{r}) = \sum_{\mu'} \alpha_{\mu'} T_{\mu'}(\mathbf{r})
\end{equation}
where
\begin{equation}\label{eq:coeff}
    \alpha_{\mu'} = \frac{(C/\tau)_{\mu'}}{\sum_{\mu^{''}}(C/\tau)_{\mu^{''}}}.
\end{equation}
In Equation~\ref{eq:coeff}, $C_\mu$ is the mode-resolved volumetric heat capacity, which also include the normalization factor arising from real-space discretization. We note that both $T_\mu(\mathbf{r})$ and $\bar{T}(\mathbf{r})$ are deviations from a reference temperature $T_0$, which we conveniently set to 0 K. Throughout this work, we refer to $\bar{T}(\mathbf{r})$ as simply the temperature. Within this formulation, the flux is 
\begin{equation}\label{eq:heat_flux}
\mathbf{J}(\mathbf{r})=\sum_\mu C_\mu\mathbf{v}_\mu T_\mu(\mathbf{r}).
\end{equation}
It is straightforward to show that after including Eq.~\ref{eq:temp} into Eq.~\ref{eq:BTE} and using Eq.~\ref{eq:heat_flux}, we obtain $\nabla \cdot \mathbf{J}(\mathbf{r}) = 0$, which is the usual steady-state continuity equation. The pore walls, denoted here by $\mathbf{r}_B$, diffusely scatter incoming phonons isotropically 
(hard-wall boundary condition), such that outgoing phonon modes $\mu'$ (i.e., those with 
$\mathbf{v}_{\mu'} \cdot \mathbf{\hat{n}} > 0$, where $\mathbf{\hat{n}}$ is the outward 
normal to the wall) satisfy
\begin{equation}\label{TBE}
T_{\mu'}(\mathbf{r}_B) =  \frac{\sum_{\mu^+} C_\mu T_\mu(\mathbf{r}_B) \mathbf{v}_{\mu^+} \cdot
\mathbf{\hat{n}}}{\sum_{\mu^+} C_{\mu^+}\mathbf{v}_{\mu^+}\cdot \mathbf{\hat{n}}}, 
\quad \mathbf{v}_{\mu'} \cdot \mathbf{\hat{n}} > 0,
\end{equation}
where $\mu^+$ refers to phonon modes incoming to the surface 
(i.e., $\mathbf{v}_{\mu^+} \cdot \mathbf{\hat{n}} < 0$) and $\mu'$ denotes the 
reflected (outgoing) phonon directions. Periodic boundary conditions are applied throughout the simulation domain, where a jump of temperature jump $\Delta T = 1K$ is also applied across the $y$-axis. The resulting boundary conditions are
\begin{eqnarray}\label{eq:periodic}
T_\mu\left(-L/2,y\right) &=& T_\mu\left(L/2,y\right),\nonumber \\
T_\mu\left(x,L/2\right) &=& T_\mu\left(x,-L/2\right) + \Delta T,
\end{eqnarray}
Since $\Delta T$ is isotropic and $\sum_\mu C_\mu\mathbf{v}_\mu =0$ is zero due to time-reversal symmetry, Eq.~\ref{eq:periodic} ensures the continuity of the heat flux across the unit-cell border. Furthermore, combining Eqs.~\ref{eq:temp}-\ref{eq:coeff} with Eq.~\ref{eq:periodic}, we have $\bar{T}(-L/2,y) = \bar{T}( L/2,y)$ and $\bar{T}(x,L/2) = \bar{T}(x,-L/2) + \Delta T$,
where we used $\sum_{\mu'^{+(-)}}\alpha_{\mu'}=1/2$. Therefore, the imposed $\Delta T$ at the phonon-mode level translates into a jump in $\bar T(\mathbf{r})$. We note that this form of periodic boundary conditions are commonly used in BTE solvers (e.g., see~\cite{hu2024giftbte})

Once Eq.~\ref{eq:BTE} is solved, the effective thermal conductivity, which is our ground truth, is computed by Fourier's law
\begin{equation}
\kappa = -\frac{1}{\Delta T}\int_{-L/2}^{L/2} \mathbf{J}(x,L/2)\cdot \mathbf{\hat{y}} dx,
\end{equation}
In absence of boundaries, the effective thermal conductivity equals the bulk one, given by 
\begin{equation}
\kappa^{\alpha\beta} = \sum_{\mu} C_\mu v_\mu^{\alpha} v_\mu^{\beta}\tau_{\mu},
\end{equation}  
which is the standard kinetic formula~\cite{ziman2001electrons}.
In this work, we choose Si as the underlying material, which gives~$\kappa_{\text{bulk}}\approx150 $W m$^{-1}$ K$^{-1}$. The second- and third-order force constants are obtained from the \texttt{AlmaBTE} database~\cite{carrete2017almabte}, which computes the forces based on the supercell approach and density functional theory~\cite{broido2007intrinsic}. The scattering times are computed on a reciprocal-space grid of $32\times 32\times 32$. Equation~\ref{eq:BTE} is solved using the open-source software OpenBTE~\cite{openbte}, which discretizes the spatial domain with the finite-volume technique, and bins the reciprocal space into a set of directional MFPs~\cite{romano2021efficient}. In this work, we employ 50 MFPs and 48 polar angles. Details about the solver implementation can be found in~\cite{openbte}

We parametrize the nanostructured material by a binary vector $\mathbf{x}$ of size 25, corresponding to a grid of 5x5 embedded pores of size 10x10 (nm) (example in Fig.~\ref{fig:data-efficiency}B), yielding a total of approximately $33$ million ($2^{25}$) possible configurations. The goal of this work is to find a surrogate for the relationship $\kappa(\mathbf{x})$. With our surrogates we aim to achieve a $5.00$\% fractional error (FE) defined as $ FE = \frac{|\kappa - \kappa_\theta |}{|\kappa|}$ with respect to the ground-truth effective thermal conductivity. The low-fidelity solver used in this study is the Fourier solver developed by the OpenBTE team (yet to be released). To enhance computational efficiency, a coarser spatial conductivity discretization is adopted: a $5\times5$ grid compared to the $100\times100$ used in the BTE solver. This resolution is the minimum required to adequately represent the $5\times 5$ pore parametrization. The Fourier solver takes as input a spatially varying thermal conductivity field, where 
the thermal conductivity within pore regions is set to zero. The governing equation reads
\begin{equation}
    \nabla \cdot \left[ \kappa(\mathbf{r}) \nabla T(\mathbf{r}) \right] = 0, \quad \mathbf{r} \in \Omega,
\end{equation}
with adiabatic boundary conditions
\begin{equation}
\kappa\nabla T(\mathbf{r})\cdot \mathbf{\hat n}=0\quad\quad\quad\quad \mathbf{r}\in \mathbf{r}_B,
\end{equation}
and periodic boundary conditions
\begin{eqnarray}
T(-L/2,y) &=& T(L/2,y)\nonumber \\
T(x,L/2) &=& T(x,-L/2) + \Delta T.
\end{eqnarray}
Similarly to the BTE case, the flux $-\kappa\nabla T(\mathbf{r})$ is continuous across the border because $\nabla \Delta T = \mathbf{0}$.

\section{Results}
\label{sec:results}
\subsection{Predictive Performance and Data Efficiency}
We compared a vanilla MLP, PEDS, and 4-model ensembles of both architectures. The training dataset was generated by randomly sampling binary 25-dimensional parametrizations and running the \texttt{openbte} mode-resolved steady-state solver on the resulting geometries; In our base experiment the training dataset was chosen to be 1000 pairs. Similarly, the validation and test sets each consist of 1000 randomly sampled pairs. No preprocessing was applied to the inputs or outputs beyond what is described below. We trained for 1000 epochs with an ADAM optimizer and a cosine schedule with maximum learning rate of $5 \cdot 10^{-3}$ and a minimum of $5 \cdot 10^{-4}$. PEDS architecture was chosen to be relatively small with 2 fully-connected hidden layers of size 64 and a resolution of 1 (5x5 grid). Consistently with previous work~\cite{Pestourie_Mroueh_Rackauckas_Das_Johnson}, smaller resolutions seemed not to yield any improvement in terms of accuracy. All activation functions are ReLU, except for the final hard-tanh layer, which constrains the output to the range $[0,\kappa_{\text{bulk}}]$. All model parameters were initialized using Xavier-normal initialization. The learnable parameter $w_\theta(x)$ controlling the geometries linear combination is the output of a fully connected perceptron with a sigmoid activation and Kaiming initialization~\cite{he2015delving}. The MLP baseline architecture was chosen to resemble PEDS', having an additional layer to match the number of parameters. Other number of layers, activation functions and initializations were tested before converging to these architectural choices. The training was parallelized and performed on 4 CPU cores. For completeness, we also included the performance of a Gaussian Process (GP) in our comparison. GPs are a widely used Bayesian approach for regression tasks~\cite{rasmussen2006gaussian}. Their flexibility and strong performance in low-data regimes make them a standard benchmark in surrogate modeling and scientific machine learning. We experimented with both RBF and Matern kernels, classic choices for modeling smooth and moderately rough functions respectively~\cite{rasmussen2006gaussian}. A Matern kernel with learned length scale and smoothness parameter, combined with epistemic white noise with inferred intensity yielded the best predictive accuracy. 

\begin{figure}[H]
    \centering
    \includegraphics[width=1.0\linewidth]{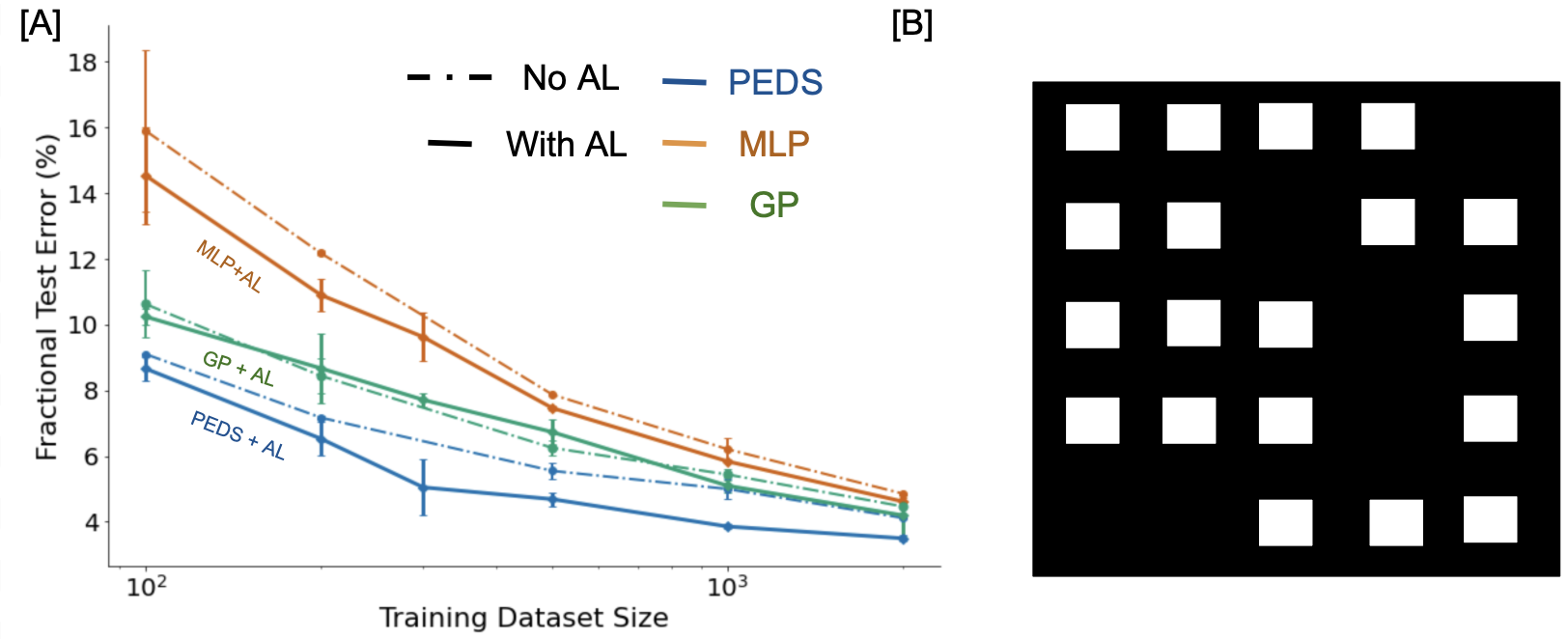}
   \caption{Panel (a) quantifies data efficiency by plotting fractional test error against training-set size for three surrogate approaches: PEDS-ENSEMBLE (blue), a MLP-ENSEMBLE (orange) and a Matern kernel GP (green), showing results both with Active Learning (solid lines) and without (dash–dot). Error bars reflect standard deviations for the 5 seeds. The main result is that physics-enhanced learning with Active Learning produces the largest reductions: PEDS paired with AL lowers the fractional error by $\sim$ 70\% relative to PEDS without AL, $\sim$75\% relative to MLP with AL, and PEDS alone already reduces error by $\sim$ 50\% compared with the MLP. The GP delivers steady and robust performance with small variance, but its improvement from AL is modest compared with PEDS. Panel (b) shows an optimized periodic pore pattern found by coupling a PEDS-ENSEMBLE surrogate and AL inside a Bayesian optimization loop to hit an effective conductivity of 30.00.}
    \label{fig:data-efficiency}
\end{figure}

\noindent To evaluate data efficiency, the experimental setup was kept fixed while varying the training dataset size. Uncertainty was quantified using a deep ensemble of four PEDS models paired with a MLP model for log-variance, all trained with the negative log-likelihood loss from Sec~\ref{sec:NLL}. The most relevant results are reported in panel (a) of figure \ref{fig:data-efficiency} and in table \ref{tab:test_error_table_AL_1}, while more extensive results are available in the Appendix. We compare fractional errors obtained by selecting a fixed number of data points through active learning. \\

\noindent The ensemble of PEDS combined with active learning reduces the fractional test error by approximately 70\% compared to PEDS without active learning and by about 75\% compared to a purely data-driven MLP trained with active learning. A single PEDS model alone lowers the error by roughly 50\% relative to a single MLP. These gains are most pronounced in the small-data regime, where PEDS starts well below data-driven baselines and reaches the 5\% error threshold with substantially fewer training samples. Across random seeds, the GP serves as a stable, low-variance surrogate and performs surprisingly well as a general-purpose baseline, though its improvement from active learning is modest compared to PEDS. Mechanistically, the strong PEDS + AL performance arises from its model structure and sampling strategy: PEDS incorporates a low-cost, physics-based Fourier core to capture the governing physics, while the neural generator learns pore-scale corrections that achieves the surrogate’s accuracy. This reduces the complexity of the learning task and, consequently, the number of expensive BTE labels required for good generalization.  AL amplifies those savings by directing costly simulations to the most informative data points. Empirically, we find the AL loop oversamples geometries in the tails of the effective-conductivity distribution. The model first captures the dominant posterior modes and then improves by incorporating these rarer, small-$\kappa$ and high-$\kappa$ examples.  In contrast, a plain MLP must learn both macro- and micro-scale behavior from data and therefore requires many more examples, and GP, while sample-efficient and uncertainty-aware, cannot exploit the physics-informed structure to the same extent. Overall these results indicate that AL delivers the biggest marginal gains when it is applied to input-space generation and paired with a physics-informed PEDS model.This effect is consistent with classical active learning surveys and with surrogate-assisted optimization practice (e.g. EGO and other related strategies)~\cite{settles2012active, forrester2007multi}. The data efficiency of PEDS is unprecedented compared to prior hybrid physics-ML surrogates ~\cite{lu2022multifidelity}. In fact, to compute the thermal conductivity, also accounting for its multi-fidelity version, the DeepONet strategy in~\cite{lu2022multifidelity} required a significantly larger number of evaluations (10x with respect to our base experiment, almost 30x if we consider our active learning). Nonetheless, these two strategies are not perfectly comparable. In principle DeepONet is a solver, and this reflects also on the choice of computing the error on the temperature field rather than on the scalar effective conductivity.

\begin{figure}[H]
    \centering
    \includegraphics[width=0.75\linewidth]{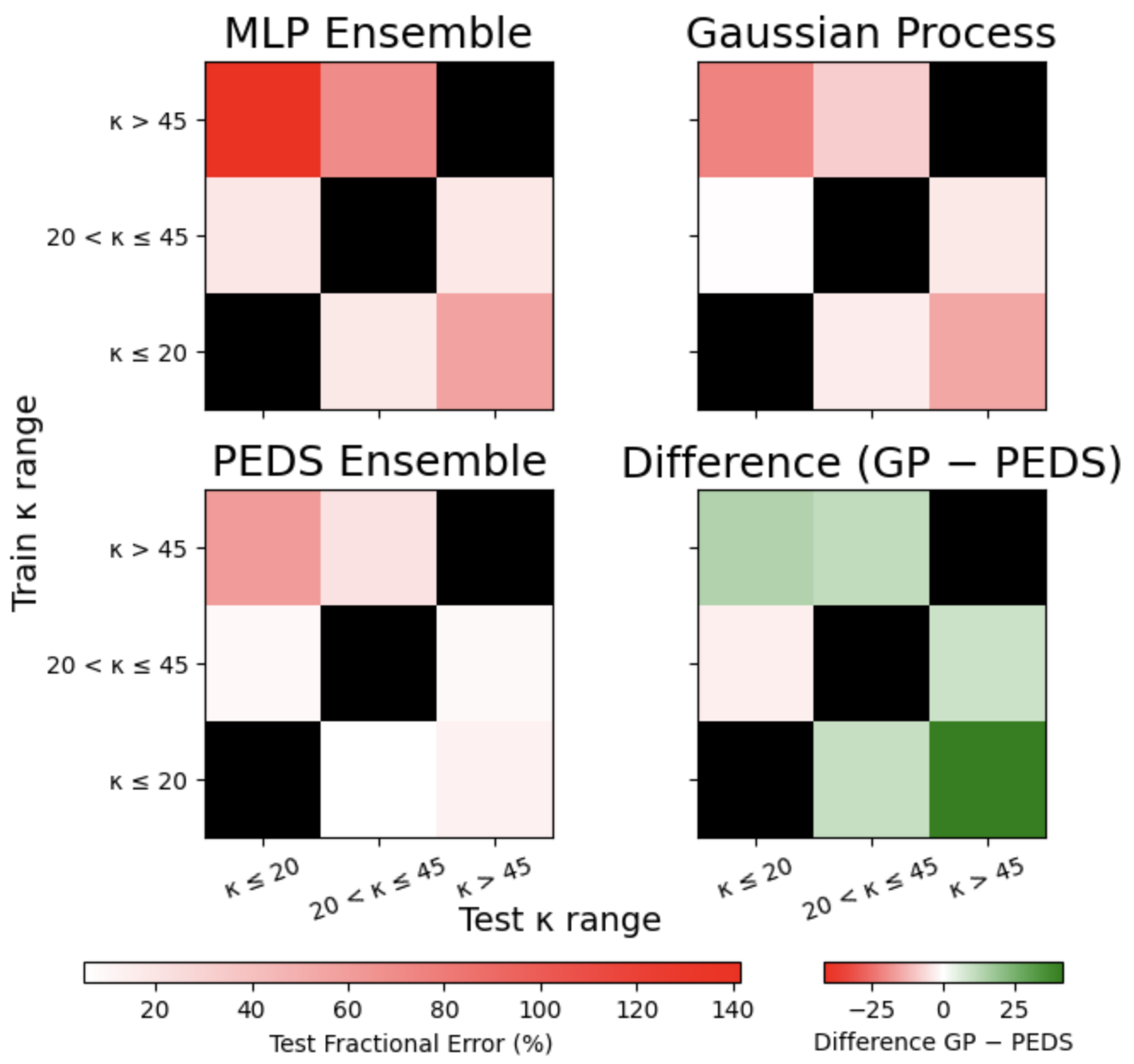}
    \caption{
Training was performed on geometries from one segment of the $\kappa$ domain, and predictions were made for geometries in other intervals. The first three matrices represent the fractional error on the test sets for the various combination of train-test splits. The last shows the comparison with the most competitive baseline, the GP (green: PEDS has smaller error; red: GP has smaller error). PEDS generalizes better than the other when tested on a out-of-distribution test set. }
    \label{fig:splits}
\end{figure}

\noindent To test the robustness of the learning to the seen data, we performed experiments on a split of the dataset. The splits correspond broadly to areas of the $\kappa$ domain associated with ballistic ($\kappa$ $\leq$ 20), diffusive ($\kappa$ $>$ 45) and the transitory phase between the two ($ 20 < \kappa \leq 45$). We train on one split of the domain and test on the other. As we can see in Figure~\ref{fig:splits}, PEDS generalizes substantially better than the other surrogates when testing it on split-distribution. Training is performed on one subset of $\kappa$ values, and predictions are made on disjoint intervals in the test set to assess out-of-distribution performance. Thanks to its physical inductive bias, PEDS is able to perform well on the whole domain when provided only with the low-$\kappa$ samples. The surrogate already encodes the diffusive behavior and therefore generalizes much better to larger $\kappa$ intervals than purely data-driven baselines. Indeed, when trained on the smallest $\kappa$ group ($\kappa$ $\leq$ 20) and tested on the mid and large $\kappa$ groups, PEDS delivers 5.1\% and 12.7\% mean fractional error respectively, dramatically lower than the MLP (17.7\%, 56.7\%) and GP (15.4\%, 54.2\%) and competitive also with MLP and GP trained on the full dataset.

\begin{table}[H]
    \centering
    \begin{tabular}{c|cccccc}
        \makecell{Training \\ Evaluations} & 
        PEDS-ENS (\%) & 
        PEDS+AL (\%) & 
        GP (\%) & 
        GP+AL (\%) & 
        MLP-ENS (\%) & 
        MLP+AL (\%) \\
        \hline
        100 & $9.10 \pm 0.01$ & $8.66 \pm 0.35$ & $10.62 \pm 1.02$ & $10.24 \pm 0.25$ & $15.89 \pm 2.46$ & $14.53 \pm 1.49$ \\
        200 & $7.17 \pm 0.06$ & $6.53 \pm 0.52$ & $8.44 \pm 0.52$ & $8.67 \pm 1.07$ & $12.17 \pm 0.08$ & $10.90 \pm 0.50$ \\
        300 & $6.72 \pm 0.15$ & $5.05 \pm 0.86$ & $7.68 \pm 0.32 $ & $7.71 \pm 0.22$ & $10.98 \pm 0.10$ & $9.62 \pm 0.74$ \\
        500 & $5.55 \pm 0.24$ & $4.69 \pm 0.20$ & $6.25 \pm 0.24$ & $6.73 \pm 0.37$ & $7.87 \pm 0.00$ & $7.46 \pm 0.07$ \\
        1000 & $5.00 \pm 0.31$ & $3.86 \pm 0.08$ & $5.44 \pm 0.17$ & $5.10 \pm 0.23$ & $6.21 \pm 0.34$ & $5.84 \pm 0.05$ \\
        2000 & $4.12 \pm 0.02$ & $3.51 \pm 0.09$ & $4.47 \pm 0.35$ & $4.20 \pm 0.65$ & $4.86 \pm 0.00$ & $4.62 \pm 0.12$ \\
    \end{tabular}
    \caption{Test error (\% mean $\pm$ std) across training evaluations for surrogates with and without Active Learning (AL). This table corresponds to the image above. Statistics are computed on 5 repetitions.}
    \label{tab:test_error_table_AL_1}
\end{table}

\subsection{Accurate and Efficient Design}
\label{sec:efficient-design}

Since currently only the single-mode BTE is differentiable on OpenBTE, direct gradient-based optimization is not readily available for our problem. As a result, we include a non-gradient-based Bayesian Optimization (BO)~\cite{Jones1998EGO, Forrester2009RecentSurrogate} strategy as a strong and widely-used baseline for optimization of expensive-to-evaluate black-box functions. The BO routine proceeds by modeling the objective as a Gaussian Process over the binary domain and selecting points that balance exploration and exploitation using an acquisition function. Our acquisition function used is the Expected Improvement (EI) \cite{Keane2006StatImprovement}.

\noindent Assuming the GP posterior at point \( \mathbf{x} \) has predictive mean \( \mu(\mathbf{x}) \) and standard deviation \( \sigma(\mathbf{x}) \), and letting \( f^* \) be the best (error minimizing) observed value so far, the EI can be computed in closed form as \( \mathrm{EI}(\mathbf{x}) = \left[ \max(f^* - f(\mathbf{x}), 0) \right] = (f^* - \mu(\mathbf{x})) \cdot \Phi\!\left(\tfrac{f^* - \mu(\mathbf{x})}{\sigma(\mathbf{x})} \right) + \sigma(\mathbf{x}) \cdot \phi\!\left(\tfrac{f^* - \mu(\mathbf{x})}{\sigma(\mathbf{x})} \right) \), where \( \Phi \) and \( \phi \) denote the cumulative distribution function (CDF) and probability density function (PDF) of the standard normal distribution, respectively. An initial Sobol (quasi-random) phase can be introduced to fill up the space, providing a warm start for the optimization encouraging exploration~\cite{snoek2012practical}. We implemented the Bayesian optimization (BO) baseline using the high-level interface of the \texttt{ax} library~\cite{ax_platform}, minimizing the $l_2$ distance to the target conductivity, $\lvert \kappa(\mathbf{x}) - \hat{\kappa} \rvert^2$, with convergence defined as reaching within 5\% fractional error of the target value. In Table~\ref{tab:design_al_error_part1} we present results for 8 example target thermal conductivities. These are non-uniformly spaced, reflecting the inherent skewness of the underlying distribution of the $30$ million candidate geometries. The figures in the table are the model error (i.e. the fractional error between the conductivity predicted by the surrogate and the actual value for that geometry) and the design error (the discrepancy, in fractional error, between the desired thermal conductivity and value for the selected geometry). The design error is the sum of model error and optimization error. PEDS+AL achieves the best average design fractional errors at $4.0\%$, dominated by the material fabrication error. The other surrogates do not perform as well with PEDS alone at an average of $8.4\%$ and GP and GP+AL at $7.4\%$ and $9.0\%$, respectively. The current best (gradient-free) optimization uses the BTE solver and reaches the average of $2.4\%$. Plain PEDS exhibits substantially higher errors than PEDS+AL, underlining the role of active learning in improving surrogate robustness in more difficult regions of the space. Notably, the total design error that is the sum of model error and optimization error is dominated by the model error. In terms of computational efficiency, the surrogates are orders of magnitude faster than BTE. More specifically, for a single optimization run with PEDS evaluation takes $0.22$ seconds (almost 3 orders of magnitude faster). However, the strong batch-independence is so that for multiple parallel optimization runs (say 128 in this case) the costs is 
0.002 (leading to 4 to 5 orders of magnitudes of improvement, 1 to 2 coming from the parallelism). Each design can be obtained within seconds to a couple of minutes, compared to several hours required for the full BTE optimization runs. Importantly, once the surrogate models are trained, they can be re-used for multiple design tasks at no additional training cost and at an overall cost comparable to solving one single BTE, amortizing the total training costs in only four optimization runs. \\

\begin{table}[H]
    \renewcommand{\arraystretch}{0.75} %
    \centering
    \begin{tabular}{@{}c|c|c|c|c|c|c@{}}
        \makecell{Kappa \\ Target} & 
        Model & 
        \makecell{Evaluations} & 
        \makecell{Computation \\ Time} &
        \makecell{Kappa \\ Optimized } & 
        \makecell{Relative \\ Model \\ Error} & 
        \makecell{Design \\ Fractional \\ Error} \\
        \hline

        \multirow{5}{*}{12.0} 
            & PEDS  & 200  & 50 s & 12.80 & 0.06  & 0.07 \\
            & PEDS+AL  & 200  & 50 s & 12.86 & 0.04  & 0.07 \\
            & GP  & 95 & $\sim$ 1s & 13.83 & 0.14  & 0.14 \\
            & GP+AL  & 111 & $\sim$ 1s & 14.58 & 0.21 & 0.21 \\
            & BTE & 106 & 318 min & 12.57 & N/A & \textbf{0.05}  \\
        \hline
        \multirow{5}{*}{15.0} 
            & PEDS  & 79  & 20s & 13.86 & 0.08 & 0.08 \\
            & PEDS+AL  & 51  & 13s & 14.56 & 0.04  & 0.03 \\
            & GP  & 58 &  $\sim$ 1s & 14.56 & 0.03  & 0.03 \\
            & GP+AL  & 54 &  $\sim$ 1s & 15.74 & 0.05 & 0.05 \\
            & BTE & 52 & 156 min & 14.88 & N/A & \textbf{0.01} \\
        \hline
        \multirow{5}{*}{20.0} 
            & PEDS  & 1  & $\sim$ 0.25 s & 18.71 & 0.06  & 0.06 \\
            & PEDS+AL  & 4  & $\sim$ 1 s & 20.31 & 0.009  & 0.02 \\
            & GP  & 12 &  $\sim$ 1s & 20.17 & 0.008 & \textbf{0.01} \\
            & GP+AL  & 4 &  $\sim$ 1ss & 20.31 & 0.009 & 0.02 \\
            & BTE & 4 & 12 min & 20.61 & N/A & 0.03 \\
        \hline
        \multirow{5}{*}{30.0} 
            & PEDS  & 100  & 25 s & 28.00 & 0.06  & 0.06 \\
            & PEDS+AL  & 8  & 2 s & 29.97 & 0.005  & \textbf{0.01} \\
            & GP  & 154  &  $\sim$ 1s & 25.50 & 0.15  & 0.15 \\
            & GP+AL  & 56 &  $\sim$ 1s & 31.57 & 0.06  & 0.05 \\
            & BTE & 8 & 24 min & 30.45 & N/A & 0.02 \\
        \hline
        \multirow{5}{*}{45.0} 
            & PEDS  & 83  & 21 s & 43.74 & 0.02  & 0.03 \\
            & PEDS+AL  & 62  & 16 s & 48.31 & 0.07 & 0.07 \\
            & GP  & 101  &  $\sim$ 1s & 40.00 & 0.11  & 0.11 \\
            & GP+AL  & 53 &  $\sim$ 1s  & 39.11 & 0.13 & 0.13 \\
            & BTE & 53 & 159 min & 42.98 & N/A & \textbf{0.01} \\
        \hline
        \multirow{5}{*}{60.0} 
            & PEDS  & 81 & 21 s & 52.8 & 0.11  & 0.11 \\
            & PEDS+AL  & 75 & 19 s & 60.10 & 0.01  & \textbf{0.01} \\
            & GP  & 96  &  $\sim$ 1s  & 57.84 & 0.04  & 0.04 \\
            & GP+AL  & 161 &  $\sim$ 1s & 52.86 & 0.12 & 0.12 \\
            & BTE & 55 & 165 min & 57.29 & N/A & \textbf{0.01} \\
        \hline
        \multirow{5}{*}{75.0} 
            & PEDS  & 84 & 21 s & 82.62 & 0.10 & 0.10 \\
            & PEDS+AL  & 113  & 29 s & 75.90 & 0.01 & \textbf{0.01} \\
            & GP  & 66 &  $\sim$ 1s & 75.65 & 0.01  & \textbf{0.01} \\
            & GP+AL  & 200 &  $\sim$ 1s & 77.79 & 0.05 & 0.04 \\
            & BTE & 66 & 198 min  & 73.59 & N/A & \textbf{0.01} \\
        \hline
        \multirow{5}{*}{85.0} 
            & PEDS  & 57  & 15 s & 98.65 & 0.16  & 0.16 \\
            & PEDS+AL  & 103  & 26 s & 93.74 & 0.10 & 0.10 \\
            & GP  & 175  &  $\sim$ 1s & 93.87 & 0.10  & 0.10 \\
            & GP+AL  & 70 &  $\sim$ 1s & 93.87 & 0.10  & 0.10 \\
            & BTE & 200 & 600 min & 80.57 & N/A & \textbf{0.05} \\
        \hline
    \end{tabular}
    \caption{Surrogates and Baseline Performances (Kappa targets 12.0–85.0). 
    The average design fractional errors are: $8.4\%$ for PEDS, $4.0\%$ for PEDS+AL, $7.4\%$ for GP, $9.0\%$ for GP+AL, and $2.4\%$ for the reference BTE solver.
    These results highlight that PEDS+AL achieves the lowest average error among surrogates, closely matching BTE while maintaining computational efficiency.}
    \label{tab:design_al_error_part1}
\end{table}

 Our main objective was to create a surrogate that enables accurate time-efficient design. In cost-benefit terms, our proposed pipeline for design tries to find a good trade-off between the high fixed costs of generating data and training the surrogate, and the high variable (per-evaluation) computational costs of optimizing directly with a black-box gradient-free method using the ground truth solver The fundamental success metric is the break-even number of designs for which choosing our strategy yields time and computation savings. 
\begin{equation}
    \underbrace{K \times N^{BTE} \times T_{BTE}}_{\text{Design Costs}} \quad \text{vs} \quad \underbrace{K \times N^{PEDS} \times T_{PEDS}}_{\text{Design Cost}} + \underbrace{T_{BTE} \times M + T_{train}}_{\text{Training Costs}}
\end{equation}
where $K$ is the number of designs, $N^{BTE}$ and $N^{PEDS}$ are the number of evaluations needed to converge to a specific effective conductivity with a Bayesian optimizer with BTE solver or with our surrogate, $M$ is the number of points generated to train the model and $T_{BTE}, T_{PEDS}, T_{train}$ are respectively the BTE and PEDS evaluation time, and PEDS fixed training time. In our case these numbers are $N^{BTE} \approx 68$, $N^{PEDS} \approx 77$, $M=300$, $T_{BTE} \approx 3\space min$, $T_{PEDS} \approx 0.22s$, $T_{train} \approx 2400s$, illustrated in Table~\ref{tab:aggregate_totals_averages}. This yields a a break-even $K \approx 4$. Additional results  where we use our surrogate with a less comparable but more parallelizable Genetic Algorithm are provided in the Appendix.

\begin{table}[ht]
\begin{tabular}{@{}lrrrrrr@{}}

Model & Tot. Evals & Avg. Evals & Tot. time (min) & Avg. Time (min) & Model error (\%) & Design error (\%) \\

PEDS       & 685 & 85.6  & 2.85   & 0.35   & 8.12 & 8.38 \\
PEDS+AL    & 616 & 77.0  & 2.56  & 0.32  & 3.55 & 4.00 \\
GP         & 757 & 94.6  & $\sim$ 0.167    & $\sim$0.0167   & 7.35 & 7.38 \\
GP+AL      & 709 & 88.6  & $\sim$ 0.167    & $\sim$0.0167   & 9.11 & 9.00 \\
BTE & 544 & 68.0  & 1632.00 & 204.00 & N/A   & 2.38 \\

\end{tabular}
\caption{Aggregate evaluations, Average evaluations, aggregate compute time in minutes, and average errors across the eight design campaigns.}
\label{tab:aggregate_totals_averages}
\end{table}

\section{Model Interpretability}
\label{sec:interpret}
\begin{figure}[H]
\centering
\includegraphics[width=1\linewidth]{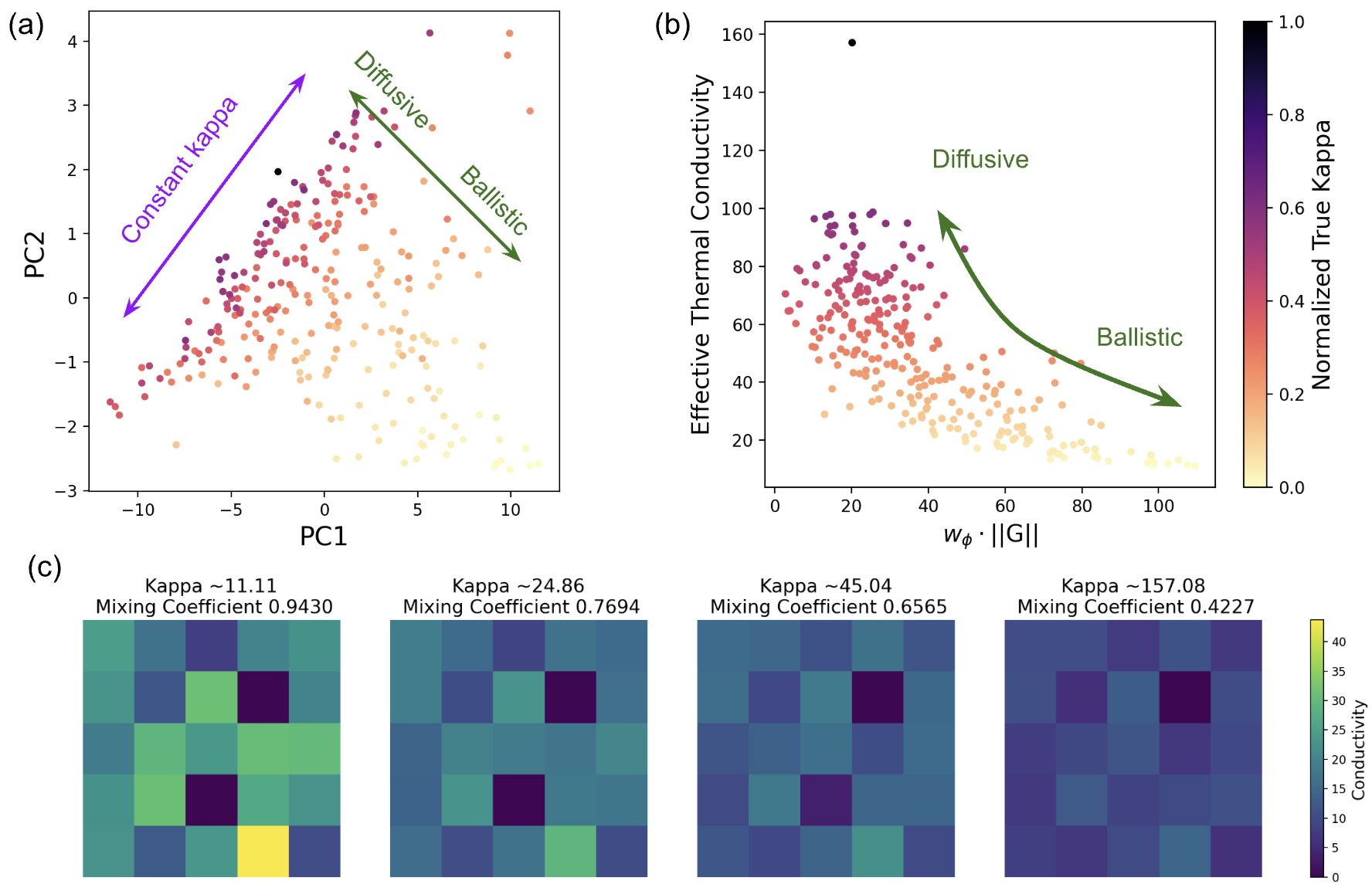}
\caption{(a) PCA of the generated conductivities projected on the first two principal components and revealing a clear trend explained by effective conductivity. (b) Relationship between the ground-truth BTE conductivity and the normalized mixing coefficient $w_\phi \|G\|$, showing that the learned parameter encodes the transition between ballistic (large $w_\phi \|G\|$) and diffusive (small $w_\phi \|G\|$) regimes. (c) Representative generated geometries, from lowest to highest conductivity, illustrating modest structural variations consistent with the low-dimensional mapping.}
\label{fig:ablation}
\end{figure}

PEDS offers a significant interpretability advantage over purely data-driven models, as the internal parameters of its low-fidelity solver inputs and outputs have clear physical meaning, enabling a direct assessment of the learning. A principal component analysis (PCA) of the 25-dimensional mode-resolved conductivity fields (5×5 grids) predicted by the neural network reveals that the structure–transport relationship is intrinsically low-dimensional: the first two principal components capture 95\% of the variance, revealing a clear trend governed by the effective conductivity (Fig.~\ref{fig:ablation}a). Two dominant modes emerge: one along which conductivity remains nearly constant, and another where it increases systematically. High-conductivity geometries occupy a wider region of the PCA space, reflecting greater structural diversity, whereas low-conductivity designs collapse into a tighter and more homogeneous cluster.
To further interpret the model’s learned physics, we examine the relationship between the ground-truth BTE conductivity and the learned mixing coefficient $w_\phi$, normalized by the generator output norm $w_\phi \|G\|$ (Fig.~\ref{fig:ablation}b). This parameter controls how much the network corrects the low-fidelity Fourier geometry. Its nonlinear dependence on $\kappa_{BTE}$ demonstrates that the model has successfully learned the transition between transport regimes: large values of $w_\phi \|G\|$ correspond to low-conductivity (ballistic-dominated) cases where the Fourier solver alone is insufficient, while smaller coefficients characterize high-conductivity (diffusive) cases adequately captured by the linear model. The structure of this mapping mirrors the nonlinear patterns previously observed between $\kappa_{BTE}$ and $\kappa_{FE}$, confirming that the learned coefficient encodes a physically meaningful transition. In this sense, we can say that PEDS  discovers the domain of validity of Fourier. Finally, representative generated geometries corresponding to low, intermediate, and high conductivities show only subtle visual differences, indicating that the latent space is smooth and effectively low-dimensional (Fig.~\ref{fig:ablation}c). The generator converges toward consistent pore configurations for each regime, supporting the hypothesis that the mapping between geometry and effective conductivity is governed by a few dominant structural modes. 

Physically, the Knudsen number (Kn) for a given geometry characterizes the transition between ballistic and diffusive transport regimes. This number effectively captures information about broadband MFP distributions in a complex geometry. Typically defined for single–MFP materials, we employ a recent computational method adapted for mode-resolved systems~\cite{hosseini2023reduced}.  For each structure, the following equality holds:

\begin{equation*}
    \frac{\kappa_{BTE}}{\kappa_{Fourier}} = \frac{1 + \text{Kn} (\ln \text{Kn} -1)}{(\text{Kn} -1)^2}
\end{equation*}

\begin{figure}[H]
    \centering
    \includegraphics[width=\linewidth]{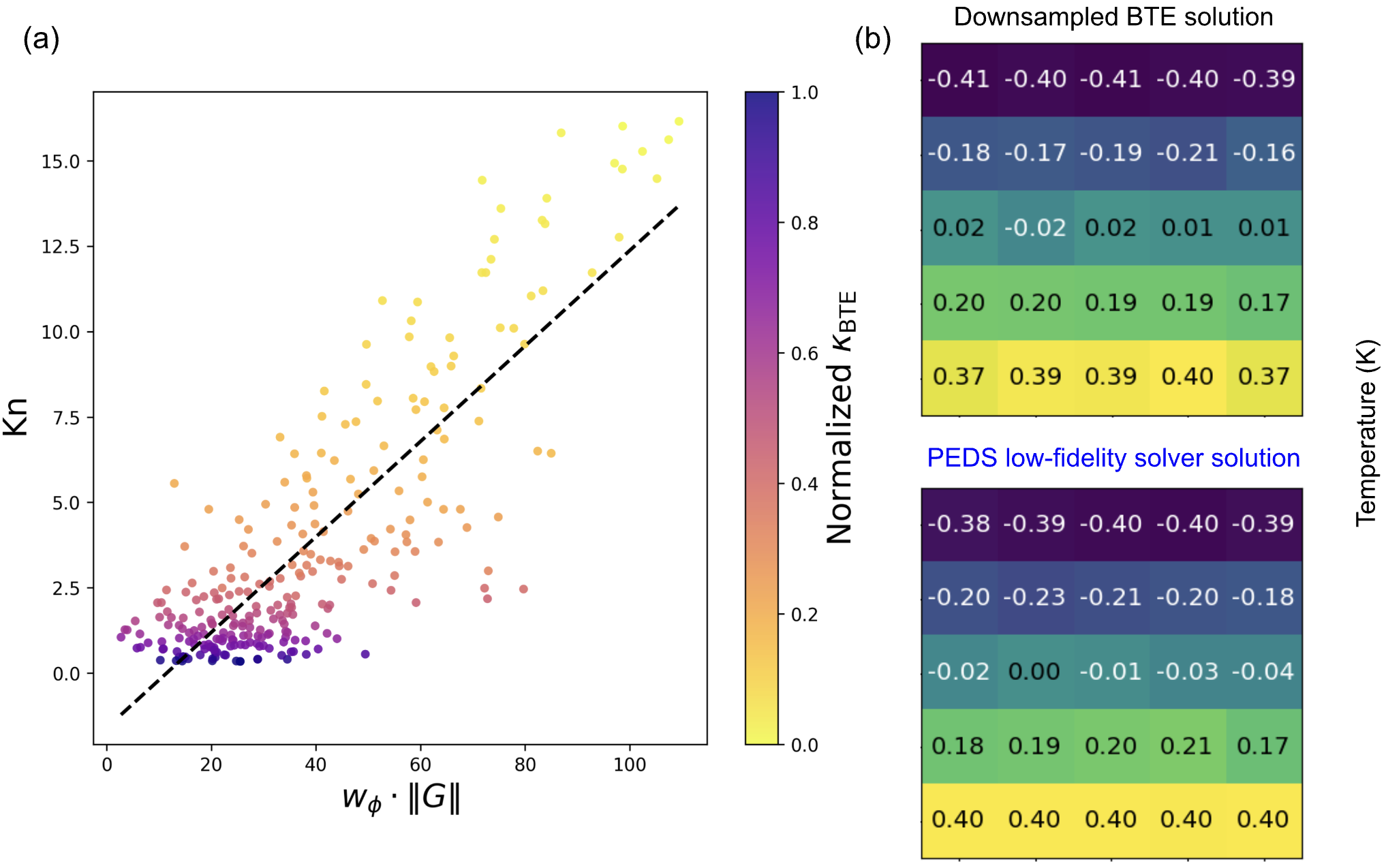}
    \caption{(a) The generated ballistic correction $w_\phi \|G\|$  is highly linearly correlated with the Knudsen number of the geometry (dashed line). The higher the Knudsen number, the bigger is the MFP compared to the representative feature size and the more important will be the ballistic correction to the diffusive model. This confirms that PEDS is able to recover the transition between diffusive and ballistic transport and compensate for the ballistic transport. (b) Qualitative comparison of the downsampled BTE temperature and the one computed internally by PEDS for the pore geometry used in Fig. ~\ref{fig:fouriererror}. This corresponds to an $8\%$ relative percentage error with respect to the BTE.}
    \label{fig:knudsen}
\end{figure}

The Knudsen number measures the ratio of the mean free path (MFP) to a characteristic geometric feature length, distinguishing structures where the MFP exceeds the feature size ($\text{Kn} > 1$, ballistic regime) from those where it is smaller ($\text{Kn} < 1$, diffusive regime). In Fig.~\ref{fig:knudsen}a, the Knudsen number is computed for a set of representative geometries and shown to correlate strongly with the generated ballistic correction $w_\phi \|G\|$. Although PEDS does not explicitly enforce similarity between coase temperature fields, it is indirectly able to reconstruct the coarse temperature fields from BTE. Using a representative set of geometries, we compare the temperature field obtained by solving the Fourier equation with the neural-corrected $5 \times 5$ thermal conductivity tensor generated by PEDS against the temperature field obtained by downsampling the corresponding BTE solution. The median percentage error is $16.89\%$, which is primarily influenced by errors along the central line, where the true temperature approaches zero and small absolute deviations result in large relative errors. To provide a more robust assessment, we also report the mean absolute error, which is $0.053\,\mathrm{K}$. When normalized by the overall temperature range (approximately $1\,\mathrm{K}$), this corresponds to a relative error of $0.05$. These results show further generalization cpabilities, highlighting the ability of PEDS to recover key interpretable features of BTE thermal transport at the solution level in addition to the effective property.

\section{Concluding remarks}
\label{sec:remarks}
Our results solidify the hypothesis, supported also by previous work~\cite{Pestourie_Mroueh_Rackauckas_Das_Johnson}, that including a simple low‐fidelity solver can impart valuable inductive bias that substantially simplifies the learning task (Fig.~\ref{fig:splits}). By removing the burden of enforcing the governing equations from the neural network, the incorporated low-fidelity physics allows it to focus only on correcting non‐diffusive phonon‐size‐effect errors. This explains its data efficiency: it inherits the low-data advantages of a physically informed baseline while avoiding the large data requirements typical of neural operator for highly mode-resolved BTE problems~\cite{kennedy2000predicting, forrester2007multi, lu2021learning, li2021fourier}. Restricting the neural network’s role to geometry transformation aligns with our view that deep parametric models excel at learning features, representations, and low-dimensional projections, while numerical solvers are best suited for capturing the underlying physics. This stands in contrast to approaches that attempt to learn full solution operators, even when only a specific property of the solution is of interest~\cite{lu2022multifidelity}.

\noindent Similarly to Ref.~\cite{lu2022multifidelity}, PEDS can be seen as a natural extension of multi‑fidelity methods, hard‑coding a cheap physics core and learning only the input correction. To our knowledge, this is the first multifidelity scientific machine learning approach that achieves accuracy for BTE using a Fourier Equation low-fidelity solver. By learning the linear combination weight parameter between macro temperature field and a nano-scale residual, PEDS inherently bridges scales, making it intuitively well suited for problems where different approaches are used for different scales. 

\noindent Future work will focus on enriching the parameterization, making it bigger and continuous, and consequently, leverage a bigger neural network. In this work, we decided to keep the same parameterization as in~\cite{lu2022multifidelity}. This relatively small number of design parameters may also help explain the competitiveness of Gaussian processes. For both GPs and neural networks, scaling to richer continuous parameterization requires exponentially more data. This phenomenon, often referred to as the curse of dimensionality, is ubiquitous in machine learning and more broadly in the computational sciences. As the number of training points increases, the computational cost of fitting standard GPs becomes prohibitive, making classical GPs computationally expensive~\cite{rasmussen2006gaussian, wang2016gaussian} and likely less attractive as the results seem to show our results. In addition, their fitting cost increases as $O(N^3)$ where $N$ is the number of training instances~\cite{lookman2019active}, while neural network training cost grows linearly with N~\cite{rasmussen2006gaussian}. 

\noindent Another practical motivation for a surrogate approach is to avoid repeated inverse design across different materials and scattering physics: Recent ML studies in phonon scattering and thermal-conductivity demonstrate that transfer learning across scattering-order approximations or material families can substantially reduce the number of expensive first-principles or BTE solves required to reach predictive accuracy~\cite{guo2023fast}; incorporating material descriptors in the learning process and applying transfer learning techniques~\cite{huisman2021survey, vettoruzzo2024advances} could allow our surrogates to extend their utility and to adapt to semiconductors other than silicon with relatively few additional high-fidelity solves~\cite{guo2023fast}. Another promising direction arises from operating in the input-space machine learning regime. Combining transferability strategies with a learned low-dimensional representation or a well-chosen set of geometric parameters can help mitigate the curse of dimensionality, enabling faster adaptation to new tasks~\cite{marzban2025hilab}.

\noindent We are currently exploring the use of the surrogate to accelerate the high-fidelity BTE solver itself, where PEDS or a neural operator can provide warm starts or learned preconditioners for iterative BTE solvers, similar to Ref.~\cite{kopanivcakova2025deeponet}. Applying these ideas to OpenBTE solvers could convert some of the cost of offline surrogate into online solver speedups and improve inverse-design throughput~\cite{li2023learningpreconditioners}. Beyond acceleration, the idea of preconditioning through learned surrogates naturally connects to transferability: a surrogate trained on one family of geometries or materials could initialize solvers for related configurations, accelerating convergence even when the underlying physics changes moderately. 

\noindent Although our approach was applied to the phonon BTE, the underlying methodology can be readily extended to the electron BTE~\cite{ziman2001electrons}, where the the low-fidelity model is the drift-diffusion model~\cite{Chen2005}. Additionally, this methodology may also extend to multiscale transport problems governed by physics analogous to the BTE. Examples include neutron transport~\cite{sun2024discrete} and rarefied gas dynamics~\cite{zhang2005lattice}, when the fluid is diluted and the continuum Navier-Stokes cannot be applied. Exploring these other applications could unveil how these surrogate models encode scale-bridging inductive biases that generalize beyond phonon transport, providing a unified framework for a broader class of problems.

\section{Acknowledgement}
We thank Steven Johnson for useful discussions. We thank Blair Yats for the artwork of PEDS. NSF SUSMED and NSF Award No.~IIS~2435905.

\section{Code and Data availability}

The code is available at \href{https://github.com/tonioeltopoquegira/PEDS_BTE}{this github repository}. The Fourier solver used for final trainings is part of a code implemented by the OpenBTE developers and not yet released. The data is available upon request.

\printbibliography

\end{document}